\documentclass{ws-mpla}

\usepackage{undertilde}

\newcommand{\be}{\begin{eqnarray}}
\newcommand{\ee}{\end{eqnarray}}

\begin{document}

\markboth{Kirill Krasnov}
{Non-metric gravity: A status report}

\catchline{}{}{}{}{}

\title{NON-METRIC GRAVITY:\\
A STATUS REPORT
}

\author{\footnotesize KIRILL KRASNOV}

\address{Mathematical Sciences, University of Nottingham, Nottingham, NG7 2RD, UK\\
kirill.krasnov@nottingham.ac.uk}

\maketitle

\pub{Received (Day Month Year)}{}

\begin{abstract}
We review the status of a certain (infinite) class of four-dimensional
generally covariant theories propagating two degrees of freedom that
are formulated without any direct mention of the metric. General relativity
itself (in its Pleba\'nski formulation) belongs to the class, so these theories
are examples of modified gravity. We summarize the current understanding of the
nature of the modification, of the renormalizability properties of these
theories, of their coupling to matter fields, and describe some of their
physical properties.

\end{abstract}

\ccode{PACS Nos.: 11.10.Gh, 04.60.-m, 04.50.Kd}

\section{Introduction}

General relativity (GR) was formulated by Einstein as a theory that dynamically
determines the spacetime metric. This theory is famously non-renormalizable,
which means that it gives only the ``low-energy'' description of gravity.
Because the Newton's constant is so small, what is low energy for gravity may
be very high energy by the standards of particle physics, so for all practical
purposes we are quite happy with the description of gravity given by GR. The
problem of quantum gravity, which is to find a satisfactory description of
gravity at ``high'' (Planckian) energy is then a very hard one as it is not
possible, and is unlikely to be possible, to probe the relevant range of
energies by direct experiments. In spite of our almost absolute ignorance as to
what happens at Planckian energy, the common consensus is that it is unlikely
that the metric description of gravity survives there. For this reason many
alternative descriptions have been developed, some of them rather radical,
e.g., even calling for abandoning of the notion of the spacetime manifold. The
most popular approach to the problem is given by string theory, which is well
defined as a quantum theory at the expense of introducing extra symmetries and
dimensions that so far are not observed.

It is not universally appreciated, however, that general relativity itself admits a
rather radical reformulation that almost entirely eliminates the spacetime metric
from the picture. This formulation was discovered a long ago by Pleba\'nski
\cite{Plebanski}, and was rediscovered more recently in \cite{CDJ} after the work
on the new Hamiltonian formulation for gravity by Ashtekar \cite{Ashtekar:1987gu}.
Ashtekar's new formulation of GR can be seen \cite{CDJ} to be just the Hamiltonian
formulation of the Pleba\'nski theory. The Plebanski formulation, together with
the idea of describing gravity by something else than the spacetime metric, leads to
a class of gravity theories much larger than GR, which is the subject of this review.

In Pleba\'nski formulation of gravity, the notion of the dynamical spacetime
metric is replaced by that of the dynamical Hodge operator. Recall that, given
a metric $g_{\mu\nu}$, the Hodge operator can be defined as the map acting on
the space of two-forms and mapping a two-form $B_{\mu\nu}$ into its ``dual''
two-form $(*B)_{\mu\nu} = (1/2)
\epsilon_{\mu\nu}^{\,\,\,\,\rho\sigma}B_{\rho\sigma}$. Here
$\epsilon_{\mu\nu\rho\sigma}$ is the volume four-form compatible with the
metric. It is easy to check that the Hodge dual operator is invariant under conformal
rescalings of the metric $g_{\mu\nu}\to\Omega g_{\mu\nu}$. The key fact is that
the converse is also true and two metrics that define the same Hodge operator
are related by a conformal transformation, see, e.g., \cite{DKS} for a simple
proof. This means that the Hodge dual operator defines a metric up to conformal
transformations, and that it can be used as the main dynamical object in a
theory of gravity, instead of the metric.

This is essentially the idea that was realized in \cite{Plebanski} (even though
it is only implicit in this work). More precisely, to describe the Hodge
operator it is sufficient to specify which two-forms are self-dual. The
self-dual forms are those satisfying $*B=\sqrt{\sigma} B$, where $\sigma=*^2$,
so $\sigma=\pm 1$ for the Euclidean and Lorentzian signatures, respectively.
Knowing the subspace $W^+$ of self-dual two-forms, the anti-self-dual forms
$*B=-\sqrt{\sigma} B$ are found as those orthogonal to self-dual ones with
respect to the scalar product in the space of two-forms given by the wedge
product (the scalar product itself depends on the choice of a volume form; the
notion of the orthogonality, however, does not). The knowledge of the subspaces
$W^{\pm}$ is equivalent to the knowledge of the Hodge operator, as is not hard
to see.

The description then proceeds as follows. One introduces three two-form fields
$B^i$. Declaring the subspace (in the space of two-forms) spanned by the fields
$B^i$ to be that of self-dual two-forms determines the metric up to a conformal
factor. This last is fixed by a choice of the volume form, for which there is a
natural choice $(vol)=(1/3)\delta^{ij}B^i\wedge B^j$, where $\delta_{ij}$ is
the Kronecker delta. The metric defined by this data is then invariant under
${\rm SO}(3)$ rotations of the triple $B^i$. This suggests that we should think
of the triple $B^i$ as of a section of a vector bundle $V\to M$ associated with
a principal ${\rm SO}(3)$ bundle over the spacetime manifold $M$. Therefore, in
addition to $B^i$, we should introduce a connection $A^i$ on $V$. It is now not
hard to write the field equations for the theory. First, it is natural to
require the connection $A^i$ to be compatible with the triple $B^i$, i.e.,
require $(D_A B)^i=0$. Given a triple $B^i$, this set of $4\times 3$ equations
can be solved for $4\times 3$ components of $A^i$ in terms of derivatives of
$B^i$. It remains to write an equation that would allow one to find $B^i$. It
is natural to require that this equation be second order in derivatives and
${\rm SO}(3)$ covariant. This points to some equation that involves $F^i$(A),
the curvature of the connection $A^i$. The obvious choice $F^i(A)=0$ does give
$6\times 3$ equations for $6\times 3$ components of $B^i$. However, the theory
that one gets this way is not very interesting as it is void of any physics
(does not have local degrees of freedom). A much more interesting choice is:
\be\label{int-1} F^i(A) = \Lambda^i_j B^j, \ee where $\Lambda^i_j$ is some
(undetermined) matrix. This equation says that the curvature is purely
self-dual as a two-form. This gives us $3\times 3$ equations $(F^i(A))_{asd}=0$
for $6\times 3$ components of $B^i$ so the theory is not well-defined as we
cannot solve for the two-form fields. One can make it into a well-defined
theory by writing an action principle that leads to (\ref{int-1}), namely:
\be\label{int-2} S[B,A,\Lambda]= \int B_i \wedge F^i(A) - \frac{1}{2}
\Lambda_{ij} B^i\wedge B^j, \ee where we raise and lower indices using
$\delta^{ij}$. Varying this action with respect to $A^i$ we get the equation
$(D_A B)^i=0$, while varying it with respect to $B^i$ gives (\ref{int-1}). Note
that only the symmetric part of $\Lambda_{ij}$ enters equation (\ref{int-2}),
so this matrix should be assumed symmetric also in (\ref{int-1}). The action
(\ref{int-2}) tells us that we should treat
$\Lambda_{ij}$ as a dynamical object and vary with respect to it. We get:
$B^i\wedge B^j=0$, which gives six equations for $B^i$. Together with
(\ref{int-1}) this gives 15 equations. Noting that three components of $B^i$
are pure gauge we thus get the right number of equations to determine $B^i$
completely. The components of the symmetric matrix $\Lambda_{ij}$ are
determined from the unused equations in (\ref{int-1}). Note, however, that the
equation $B^i\wedge B^j=0$ implies, in particular, that the volume form $(vol)$
constructed above is zero. Also, it is rather obvious that there is no Hodge
dual operator for which the space $W^+$ of self-dual two-forms is null,
$W^+\wedge W^+=0$. So, this theory is not of any physical
interest.\footnote{Its Hamiltonian analysis shows that this is simply GR with
the Hamiltonian constraint removed.}

To get an interesting theory, we will impose one constraint on the symmetric
matrix $\Lambda_{ij}$ that appears on the right-hand-side of (\ref{int-1}). If
we do so, then only five of the six components of $\Lambda_{ij}$ remain
independent. Equation (\ref{int-1}) then gives ten equations for the components
of $B^i$ (this is the number of equations in GR) plus five equations for the
independent components of $\Lambda_{ij}$ in terms of the second derivatives of
$B^i$. Variation of the action (\ref{int-2}) with respect to the independent
components of $\Lambda_{ij}$ gives an additional five equations on $B^i$, which
gives just enough equations to determine the two-form fields.

What constraint can be imposed on $\Lambda_{ij}$? It is clear that the five
independent components of $\Lambda_{ij}$ can only be chosen to be those of its
{\it traceless\/} part, which we denote by $\Lambda_{ij}-(1/3)\delta_{ij} {\rm
Tr}(\Lambda):=\Psi_{ij}$. Indeed, this is the part that forms an irreducible
representation with respect to the action of ${\rm SO}(3)$. The trace part
${\rm Tr}(\Lambda)$ then becomes a function of its traceless part $\Psi_{ij}$
that we denote by $-\phi(\Psi)$ (the minus sign is to agree with certain
conventions; see below). Thus, we have: \be\label{lambda} \Lambda_{ij} =
\Psi_{ij} - \frac{1}{3} \delta_{ij} \phi(\Psi). \ee

The theory defined by the action (\ref{int-2}) where only the traceless part
$\Psi_{ij}$ of $\Lambda_{ij}$ is an independent variable (\ref{lambda}) gives
the right number of equations to solve for the components of the two-form field
$B^i$ and thus determine the Hodge dual operator, which in turn (together with
the volume form) determines the metric. We also note that the volume form
$(1/3)\delta_{ij}B^i\wedge B^j$ is no longer trivial: the equation one obtains
by varying the action (\ref{int-2}) with respect to $\Psi_{ij}$,
\be\label{metricity} B^i\wedge B^j - \frac{1}{3} \delta^{ij} (B^k\wedge B_k) =
\frac{1}{3} \frac{\partial\phi}{\partial \Psi_{ij}} (B^k\wedge B_k), \ee no
longer implies any relation for the volume form. We thus get a theory with
potentially interesting physical implications. It becomes even more interesting
after one observes that, for $\phi(\Psi)=\Lambda=\mbox{const}$, the above
theory is nothing else but the Einstein's GR in disguise (i.e., in its
Pleba\'nski formulation), with $\Lambda$ being the cosmological constant (our
choice of the sign in (\ref{lambda}) was motivated precisely by the desire to
agree with the usual convention for the sign of $\Lambda$). Therefore, what we
have obtained is an infinite family of generalizations of general relativity,
where to fix a theory one has to specify the function $\phi(\Psi)=\phi({\rm
Tr}(\Psi^2),{\rm Tr}(\Psi^3))$ of two invariants that can be constructed from
the traceless matrix $\Psi$. The above counting of the field equations suggests
that this theory makes sense for an arbitrary $\phi(\Psi)$. A more detailed
analysis, whose results we will review in the next section, shows that this is
indeed the case.

Our above discussion was motivated by the idea to describe gravity as a
dynamical theory of an object other than the spacetime metric. We have seen how
GR itself can be reformulated in these terms, and how an infinite class of
gravity theories different from GR and parametrized by a function $\phi(\Psi)$
is obtained this way. In the next section we will review the (classical)
properties of this class of theories without worrying too much about the
principle that can fix $\phi$. We will see that, for an arbitrary $\phi$, the
gravity theory (\ref{int-2}) resembles GR in many ways. Still, there are some
new physical effects predicted by this theory. It is important to emphasize
that, at the level of the classical physics, nothing forces us to depart from
the familiar ground of GR, and the new class of theories described above may be
viewed as not more than a mathematical curiosity. However, when one considers
the quantum theory of usual GR in Pleba\'nski (Hodge operator) formulation, one
{\it is\/} forced to consider theories more general than GR, similar to what
happens in quantum gravity in the metric formulation. In the usual metric
formulation of quantum gravity, one has to introduce theories with higher
derivatives and, thus, with rather unpleasant properties (instability etc.). In
the Hodge operator formulation of gravity, many (if not all; see Section
\ref{sec:quantum}) terms that have to be added to the action do not change the
character of the theory so drastically, i.e., do not introduce higher
derivatives and new DOF. The quantum corrected theory is one of the class
(\ref{int-2}) with (\ref{lambda}). Thus, one does have to take this class of
gravity theories seriously, as the modifications they describe will be induced
by quantum corrections. It is a different question whether any of these quantum
corrections can survive and be of relevance at low (astrophysical) energy
scales. It can only be answered after the quantum mechanical behavior of this
class of theories is understood. But even in the absence of such understanding,
it is interesting to quest what kind of new physical effects can be expected.

The theory (\ref{int-2}) with (\ref{lambda}) was first proposed in
\cite{Krasnov:2006du}, where renormalizability properties of GR in Pleba\'nski
formulation were considered. It was only later appreciated that it is also
quite interesting as a purely classical theory, the point of view which was
developed in \cite{Krasnov:2007uu}, \cite{Krasnov:2007ky}. The Hamiltonian
analysis of this theory (in a slightly different version) was done in \cite{Bengtsson:2007zx}
and more recently in \cite{Krasnov-recent}. We will describe the main results of the classical
analysis in the next section. The current understanding of the quantum
mechanical behavior of this class of theories is reviewed in Section
\ref{sec:quantum}.

\section{Classical properties}
\label{sec:classical}

\subsection{Hamiltonian formulation}

The canonical description of the class of theories introduced above is easily
obtained. This was done in the ``pure connection'' formulation in
\cite{Bengtsson:2007zx} and starting directly from (\ref{int-2}) in
a more recent work \cite{Krasnov-recent}. One finds,
exactly like in the usual GR--Pleba\'nski--Ashtekar case, that the phase space
is parametrized by the canonically conjugate pairs
$(A_a^i,\tilde{\sigma}^{ai})$, where $A_a^i$ is the spatial component of the
connection, and $\tilde{\sigma}^{ai}$ is the momentum (tilde denotes the
density weight). The theory is fully constrained (no Hamiltonian). The
constraints are: the Gauss constraint $D_a \tilde{\sigma}^{ai} = 0$, where
$D_a$ is the spatial covariant derivative with respect to $A_a^i$; the
diffeomorphism constraint $\tilde{\sigma}^{ai} F_{ab}^i=0$, which both take
exactly the same form as they do in Pleba\'nski theory (or the new Hamiltonian
formulation \cite{Ashtekar:1987gu}); and, finally, the Hamiltonian constraint
that becomes: \be\label{ham} \epsilon^{ijk} F_{ab}^i \tilde{\sigma}^{aj}
\tilde{\sigma}^{bk} + \frac{1}{3} \utilde{\epsilon}_{abc}\epsilon^{ijk}
\tilde{\sigma}^{ai} \tilde{\sigma}^{bj}\tilde{\sigma}^{ck}\, \phi\left(
(F_{ab}^{(i} \epsilon^{j)kl} \sigma^{ak} \sigma^{bl})_{\rm tr-free}
 \right)=0,
\ee where the trace-free part of the matrix in the argument of the function
$\phi$ is taken. Note that the momentum variables used in the argument of
$\phi$ have no density weight. For $\phi=\Lambda=\mbox{const}$, the above
Hamiltonian constraint is just that of the Ashtekar Hamiltonian formulation
\cite{Ashtekar:1987gu} of GR with the cosmological constant.

The constraints described are first class, see \cite{Krasnov-recent} for a verification of 
this. Counting of the degrees of freedom (DOF) is then exactly the same as in GR: we have $3\times 3$
configurational DOF, minus three Gauss constraints, minus three diffeomorphism
constraints, minus one Hamiltonian constraint, which gives two propagating
degrees of freedom. The class of theories in question is thus an infinite
(parametrized by a function of two variables) class of four-dimensional
generally covariant generalizations of GR propagating two degrees of freedom.

\subsection{``Curvature'' dependent cosmological ``constant''}

We have just seen that, at the level of the canonical formulation, the only
modification as compared with the usual GR case is that the cosmological
constant gets replaced by a non-trivial (and arbitrary) function $\phi$ of the
``curvature'' $(F_{ab}^{(i} \epsilon^{j)kl} \sigma^{ak} \sigma^{bl})_{\rm
tr-free}$. In GR--Pleba\'nski theory this symmetric traceless tensor is nothing
else but the Weyl part of the Riemann curvature tensor. For this reason, we
will continue to refer to this tensor as ``curvature'' for any theory of the
class under consideration. Note that this tensor has dimension of $1/L^2$, $L$
being length, as is appropriate for the curvature.

The main physical implication of this modification is that the cosmological
constant observed in the regions of relatively low curvature does not have to
be the same as that observed in the regions of relatively high curvature. In
particular, the cosmological constant that appears in the Friedmann equations
governing the evolution of a homogeneous isotropic Universe is given by
$\lambda_{\rm cosmological}=\phi(0)$ (the argument of the function $\phi$ in
this case is zero due to symmetries) and does not have to be equal to the large
curvature effective cosmological constant at Planck scales. This gives a
possible mechanism for solving the ``cosmological-constant problem'', which is
to explain why the observed cosmological constant is so different from the
cosmological constant of the order $1/l_p^2$, where $l_p$ is the Planck length,
expected to be induced by the Planck-scale physics. The cosmological constant
induced at Planck scales $\phi( 1 / l_p^2 )$ (in the regime of extremely high
curvatures $1/l_p^2$) may well be of the order of $1/l_p^2$, but this would not
have any observable effect provided the value $\phi(0)$ of the function $\phi$
at zero curvature is small. The challenge is then to explain what type of
physics fixes the form of the function $\phi$, and show that the physical
$\phi$ indeed has the properties required. We will return to this (open)
question in the next section.

\subsection{Homogeneous isotropic cosmology}

As we have already mentioned, in the case studied by Friedmann, due to high
symmetry, the ``curvature'' tensor in the argument of the function $\phi$
vanishes, and predictions of the theory with non-trivial $\phi$ are the same as
those of GR with the cosmological constant $\phi(0)$. Thus, the Friedmann
equations do not get modified. However, the theory of cosmological
perturbations does get modified. Work is currently in progress to study these
modifications.

\subsection{New physical effects in the spherically symmetric case}

The spherically symmetric problem for the theory in question was solved in
\cite{Krasnov:2007ky}. It was shown that the class of theories under study
admits an analog of the Birkhoff's theorem: a spherically symmetric solution is
necessarily static. Importantly, in this case there is just one invariant that
can be constructed from the matrix $\Psi: {\rm Tr}(\Psi)^2\sim\beta^2, {\rm Tr}(\Psi)^3\sim\beta^3$,
so the function $\phi$ becomes that of a single argument $\phi=\phi(\beta)$. 
To describe the main new physical
effects that appear in the case of a non-trivial function $\phi$, let us assume
that this function has the form of a step-function, taking one, approximately
constant value in the region of relatively large curvatures and another
constant value in the region of relatively small curvatures. Thus, we are
envisaging a scenario in which the function $\phi$ defines a scale, which is
the curvature scale at which the change takes effect. In other words, the
curvature scale is defined as the value of $\beta$ for which the dimensionless
quantity $\partial\phi/\partial\beta$ is significantly different from zero.

Assuming that the length scale defined by $\phi$ is sufficiently large (as
compared to the horizon radius of a central body), close to the horizon we will
have the region of ``large'' curvatures described by a (dS) Schwarzschild
solution. As one goes further away from the horizon, the curvature $\beta$
starts decreasing as $\beta\sim r_s/r^3$ ($r_s$ is the Schwarzschild radius)
and one will eventually enter the region in space (and in the curvature space)
where the function $\phi$ starts to change. One finds that when the modulus of
$\partial\phi/\partial\beta$ is nowhere of the order of unity the curvature
$\beta$ (which no longer has the interpretation of the Weyl curvature)
continues to decrease, and one eventually enters the other region of constant
$\phi$. In this region the solution is again (dS) Schwarzschild, but with, in
general, a different value of the cosmological constant. The main new physical
effects, apart from the changing cosmological constant, are: (i) The observed
value of the mass of the spherically symmetric object is different in the
``high'' and ``low'' curvature regions; (ii) There is an additional redshift
occurring as compared to the case of the usual Schwarzschild. These effects are
described by the following simple formulae: \be
\frac{r_s(\beta_1)}{r_s(\beta_2)} = Z(\beta_1,\beta_2), \qquad
\frac{f(\beta_2)}{f(\beta_1)}=\frac{g(\beta_1)}{g(\beta_2)}Z(\beta_1,\beta_2),
\ee where $f^2$ is the {\small 00} component of the metric,
$g^{-2}(r)=1-(r_s/r)-(1/3)\phi r^2$ is its inverse $rr$ component, as usual for
a (dS) Schwarzschild solution and \be\label{redshift} Z(\beta_1,\beta_2) =
e^{\int_{\beta_1}^{\beta_2} \frac{\phi_\beta}{6\beta} d\beta} \ee is the
redshift factor. Note that we are using the sign convention for $\phi$
different from that in \cite{Krasnov:2007ky}.

To describe the effect of mass ``renormalization'' in words, let us assume that
the function $\phi$ increases with curvature, which is consistent with the
assumption that $\phi$ is large at Planck scale curvatures and very small at
zero curvature. Then if we take for $\beta_{1,2}$, $\beta_1<\beta_2$ the
characteristic values for ``small'' (far away from the object) and ``large''
(close to the body) curvatures, respectively, the ``redshift'' factor is
greater than one, which means that the apparent gravitating mass of the object
{\it increases\/} as one moves further away from it. This simple observation
may be able to explain the phenomenon of missing mass (``dark matter'') as a
purely gravitational effect. Some order-of-magnitude
estimates as to what the relevant curvature scale $l$ must be for such an
explanation to be possible are given in \cite{Krasnov:2007ky}. 

The reader should keep in mind that the above discussion, as well as that in
\cite{Krasnov:2007ky}, are just first attempts to extract physical predictions
from the new theory, and that a concrete model for ``dark matter'', together
with a mechanism that would fix the form of the function $\phi$, is yet to be
proposed. We find it very encouraging, however, that the direction in which the
mass of a spherically symmetric object gets ``renormalized'' (under the
assumption of a function $\phi$ increasing with curvature) is not in conflict
with the phenomenon of missing mass.

The other effect that the modified gravity theory predicts is that of an
additional redshift (blueshift). The above formula tells us that when
light is emitted in the region of ``high'' curvatures and travels away from
the body into the region of ``low'' curvatures, in addition to the usual
relativistic redshift, there will be (under the assumption that $\phi$ is
an increasing function of the curvature) an additional blueshift that will
act in the opposite direction, increasing the photon's energy. If, however,
light travels from the regions of ``low'' curvature to those of ``high''
curvature close to the body, then there will be an
additional redshift effect. As is discussed in \cite{Krasnov:2007ky},
this effect may be of astrophysical significance.

\subsection{Avoidance of the black-hole singularity}

Yet an additional reason to take the class of theories described
seriously is that the behavior of all fields inside a black hole is much less dramatic than in GR.
Let us, as before, assume that the function $\phi(\beta)$ increases
as the curvature increases, e.g. becomes of order $1/l_p^2$ at
Planckian curvatures. The derivative $\phi_\beta$ of $\phi$ would also
normally grow. Then somewhere inside the black hole a point will be reached where
$\phi_\beta$ becomes $1/3$ (this value is due to the chosen coefficient in front of
$\phi$ in (\ref{lambda})). As a detailed analysis of \cite{Krasnov:2007ky} shows, at this point
the metric to be constructed from the two-forms $B^i$ becomes no longer
defined (singular). However, all the dynamical fields of the theory,
namely both $B^i$ and $A^i$, remain finite. Thus, one can evolve through
this surface and enter a new region, where the metric again becomes
defined, and which is absent in the Schwarzschild solution. The resulting
conformal diagram is given in \cite{Krasnov:2007ky}. Thus, the
theory no longer ``carries the seeds of its own destruction'', something
of great importance for its logical consistency. Of course, the removal
of the singularity in the spherically symmetric solution
is not equivalent to the absence of singularities in all possible situations,
but this result possibly indicates a much stronger ``avoidance of
singularity'' property of the theory.

\subsection{Coupling to matter}

Before one can take the theory described seriously one must make
sure that matter degrees of freedom can be coupled to it. In general relativity
this is straightforward, as matter couples directly to the metric whose
dynamics the theory describes. From the
construction of our theory as it was presented in the Introduction
it was clear that it is the Hodge operator that should treated as
the fundamental object of the theory, not the metric defined by it.
It is thus a crucial question if matter fields can be consistently
coupled directly to our basic dynamical fields, which are the triple of
two-forms $B^i$ and the connection $A^i$. For gauge fields,
this task is easy, as it is exactly the Hodge operator, not the metric,
that is required to write down the Yang-Mills (YM) action. The action
that couples the YM gauge field directly to $B^i$ reads \cite{CDJ}:
\be
S_{YM}[a,\varphi]=\int \varphi_i^\alpha B^i \wedge f^\alpha(a) - \frac{1}{2}
\varphi_i^\alpha \varphi_j^\alpha B^i \wedge B^j,
\ee
where $\alpha,\beta$ are Lie algebra indices for the gauge group in
question, $\varphi_i^\alpha$ is an auxiliary field, similar to $\Psi_{ij}$ in the
case of gravity, and $f^\alpha(a)$ is the curvature of the YM connection $a^\alpha$.
It is not hard to show that, in the geometric optics approximation,
this theory describes massless quanta propagating along the null
geodesics of the metric defined by $B^i$. This holds for a general
two-form field, independently of whether $B^i$ satisfies Pleba\'nski
or generalized field equations. Thus, it is encouraging that the
gauge fields can be coupled to the class of gravity theories
under consideration so seamlessly.

Coupling to other fundamental fields whose existence we know for sure ---
fermions --- is a much more tricky business. The action proposed for this
purpose in \cite{CDJ} {\it does not extend\/} beyond the case of GR. The reason
for this is that the field equations that follow from the action in \cite{CDJ}
form an over-constrained system, and this system is only consistent when the
triple $B^i$ satisfies equations (\ref{metricity}) with zero right-hand-side,
i.e., when the gravity theory is GR. For a theory with general $\phi$ the
combined theory gravity + fermions with the action proposed in \cite{CDJ} is
simply inconsistent. One therefore has to look for a different description of
fermions. Work on this important issue is currently in progress.

\section{Renormalization}
\label{sec:quantum}

\subsection{Counterterms}

Renormalizability properties of general relativity in Pleba\'nski formulation
were considered in \cite{Krasnov:2006du}. Simple power counting arguments show
that an infinite set of counterterms must be added to the action, and that in
this sense the theory is non-renormalizable, as expected. However, in
Pleba\'nski formulation, there is more field redefinition freedom than in GR.
After using the available field redefinitions one easily shows that there is
one infinite set of counterterms that gets combined into the function
$\phi(\Psi)$ that we introduced in (\ref{lambda}), and another (also infinite)
set of counterterms containing covariant derivatives of $\Psi_{ij}$. These
other terms, not considered in the theory (\ref{int-2}), are, schematically, of
the form $\Psi \ldots \Psi (D\Psi)^4$, $\Psi \ldots \Psi (D\Psi)^2 B$, $\Psi
\ldots \Psi (D\Psi)^2 F$, where $\Psi \ldots \Psi$ stands for a product of the
matrices $\Psi_{ij}$. With the addition of such terms to the action, the field
equations become higher order in derivatives and the (relative) simplicity of
the theories (\ref{int-1}) is lost. This is reminiscent of what happens in the
usual metric-based formulation, where one is forced to add to the action
counterterms that lead to higher-derivative field equations. The main
difference between the two cases is that, in the Hodge operator based theory,
one can incorporate an infinite number of quantum corrections into the action
{\it without\/} significantly changing the properties of the theory, while in
the metric based gravity even the simplest quantum induced modification has
rather dramatic consequences.

It is clear that quantum corrections described by the action (\ref{int-2})
can be expressed in the purely metric formulation. Indeed, we have described in
the Introduction how the theory (\ref{int-2}) describes the metric of the spacetime,
albeit implicitly. Thus, for a given $\phi$, field equations can be, at
least in principle, expressed as field equations for the metric tensor. It is
clear that (for a non-constant $\phi$) one will get equations involving higher derivatives of the metric.
It is also clear that an infinite expansion in derivatives must be present,
as no finite number of derivatives would lead to a theory with just two
propagating degrees of freedom (higher derivatives lead to higher number
of DOF as they require more initial data). Thus, the theory (\ref{int-2})
describes an infinitely large class of purely metric theories with infinite
number of derivatives, where the sums in derivatives have been re-summed
to get a theory with two propagating DOF. Thus, the class of theories under
consideration does describe the usual quantum corrected metric gravity.
What is unusual is the parametrization of these quantum corrections.

For a more general class of theories, namely those containing terms involving
the derivatives of $\Psi_{ij}$, a relation to the purely metric formulation
is not so clear. However, it can be expected that they also correspond
to GR plus an infinite set of higher derivative terms, that set
being more general than the one that arises from (\ref{int-2}).

\subsection{Renormalizability conjecture}

The main open problem is whether the class of quantum corrections described by
(\ref{int-2}) is complete. In other words, the main question is whether the
terms containing the derivatives of $\Psi_{ij}$ that can be added to the action
on dimensional grounds are indeed necessary as counterterms if one starts from
a theory that does not contain such terms. Our usual experience with quantum
fields suggests that they are: what can be added to the action should be added.
However, there are some rather strong indications that make the present author
to believe that the theories (\ref{int-2}) are different in this respect. 

First, the author considered a
``gauge fixed version'' of the Pleba\'nski theory with the action given by:
\be\label{gauge-fixed} S_{g.f.} = \frac{1}{2} \int (\Lambda^{-1})_{ij}
F^{+i}(A) \wedge F^{+j}(A), \ee where a background metric is chosen and
$F^{+i}$ is the self-dual part of the two-form field strength $F^i$ in this
metric, and $\Lambda_{ij}$ is given by (\ref{lambda}). The above expression
gives the ``bosonic'' part of the action, which also contains some ghost and
gauge-fixing terms. Apart from the presence in (\ref{gauge-fixed}) of a
non-trivial prefactor $(\Lambda^{-1})_{ij}$ the theory studied by the author is
the same as the so-called Donaldson theory; see
\cite{Labastida:1988qb,Birmingham:1991ty}. The action (\ref{gauge-fixed}),
together with the other (ghost) terms can be obtained from (\ref{int-2}) by a
formal gauge fixing procedure. The theory (\ref{gauge-fixed}) can be shown to
behave rather nicely under the renormalization. In particular, for this theory
no terms containing derivatives of $\Lambda_{ij}$ arise as
counterterms. This was reported in \cite{Krasnov:2006du}, where it was also
conjectured that the same property holds for the class of theories
(\ref{int-2}). The ``gauge fixing'' procedure that leads from the gravity
theory of interest to (\ref{gauge-fixed}) is only understood by the author at
the formal level, this is why a detailed derivation of these results is still
unpublished. A completely satisfactory treatment would include answering the
question how the graviton degrees of freedom are represented by the gauge fixed
action, a question currently beyond the author's understanding. Thus, the
``renormalizability'' property of the class of theories (\ref{int-2}), namely
the property that it is closed under the renormalization, remains a conjecture.

Some further indications in support of the conjecture have started to emerge
more recently. The conjecture can only be true if the class of theories
(\ref{int-2}) possesses some (presumably hidden) symmetry that prevents the
terms containing derivatives of $\Psi_{ij}$ from appearing. If this is so, what
can this symmetry be? It has recently been realized by the author that one way
to view the theories (\ref{int-2}) is to regard them as the topological BF
theory (whose action is given by the first term in (\ref{int-2})) in which a
part of the topological symmetry is {\it gauge fixed} by the second
$\Lambda_{ij}$ term. Different amount of gauge fixing leads to different
theories. Thus, e.g., the theory (\ref{int-2}) with all components of
$\Lambda_{ij}$ considered independent removes more gauge
symmetries of BF theory than it is necessary to get GR and leads to an uninteresting model. If one
accepts this point of view seriously, it then starts to look like the true
gravity theory is given by the topological BF theory ``gauge fixed'' by some
mechanism to be understood. We are then seeing gravity simply because we have
means of asking the topological BF theory non gauge-invariant questions. This
can only be the case if we have fields coupling to our theory in a non-gauge
invariant way, which is indeed the case: the coupling of matter fields to the
two-form fields $B^i$ breaks the topological symmetry of BF theory. This points
in the direction of matter fields as being responsible for the ``gauge fixing''
that leads to gravity and for an ``illusion'' of purely gravitational degrees
of freedom. Needless to say, these ideas are in their very preliminary stages
of development. But they point out in the direction of the topological symmetry
of BF theory as being behind the scene in the theories of gravity under
consideration. In the opinion of the author, it is a better understanding of
all these issues that will one day help to establish the true status of the
``renormalizability'' conjecture.

\subsection{If the conjecture is true}

This is the most optimistic scenario which does not conflict with anything
known. The renormalizability property would give us a window into the Planck
scale physics, as we could then conclude that the gravity theory at the Planck
scale is a (quantum) theory from the class (\ref{int-2}), with some function
$\phi$ that needs to be determined. In spite of our ignorance as to the form of
this function, the above statement would be extremely strong, as it would give
us means to make at least qualitative statements about the Planck-scale
physics.

What about the form of the function $\phi$? If the conjecture is true, then the
renormalization group flow for theories (\ref{int-2}) is a flow in the
(infinite dimensional) space of functions $\phi$. The best-case scenario is
then that of the asymptotic safety of Weinberg \cite{Weinberg}, which is that
this renormalization group flow has a non-trivial and reasonably well-behaved
ultra-violet fixed point. It would then make sense to use the fixed point
theory as the gravitational theory singled out by its extremely appealing
renormalizability properties. This is one possible scenario for fixing the form
of the function $\phi$. Note, however, that the coupling of gravity to matter
will most probably have a strong effect on its UV behavior, so it would be
unreasonable to study this question in the domain of pure gravity. But the idea
of fixing the form of the function $\phi$ to be that at the UV fixed point can
as well be used when matter is present. This gives at least a preliminary
scenario. Clearly much more work is required before a ``correct'' scenario for
fixing the form of $\phi$ is found.

\subsection{If the conjecture is false}

This is what one's experience with quantum fields in Minkowski spacetime would
suggests. If this is the case, one is back where one started: quantum gravity
is non-renormalizable, and its quantization does not seem to give any insight
on how it behaves at Planckian energies. Nevertheless, the class of theories
described in the present review may be of interest even in this case. As we
have described above, the theories (\ref{int-2}) do incorporate at least some
of the quantum corrections to GR. One may then argue that, at ``low'' energies,
the terms containing the derivatives of $\Psi_{ij}$ are less important and that
the low energy limit of the quantum corrected theory of gravity is given by
(\ref{int-2}). This is similar to what we know to happen with the usual
metric-based gravity: the terms containing higher derivatives become
insignificant at low energies and drop out. It would still be a challenge to
explain how a non-trivial function $\phi$ in (\ref{lambda}), if viewed as a
result of quantum corrections to GR, could survive at low energies, but this is
not impossible, in case, for example, when there is a low energy scale present
in the matter Lagrangian. All in all, even in case the renormalizability
conjecture fails, the class of theories (\ref{int-2}) is of interest as a very
simple class of ``modified gravity'' theories that does not contain new degrees
of freedom.

\section{Conclusions}

A class of gravitational theories propagating two degrees of freedom and
formulated without any direct reference to the spacetime metric was described.
A theory from this class is specified by an (arbitrary) function $\phi$ of two
arguments. General relativity belongs to the class considered and corresponds
to $\phi=\mbox{const}=\Lambda$, $\Lambda$ being the cosmological constant. One
way to describe the nature of the modification is to say that the cosmological
constant became a non-trivial function of the ``curvature''. However, the
theories we consider cannot be obtained by simply inserting a scalar function
of the Weyl curvature into the Lagrangian. The construction is much more
subtle, and involves, in particular, replacing the paradigm of gravity being a
dynamical theory of the metric by a new paradigm of gravity being about the
dynamics of the Hodge dual operator. We have described some simple physical
consequences of this modification of gravity, as well as reviewed the status of
the question of coupling these theories to matter. Thus, gauge fields couple
seamlessly, but no descriptions for fermions in this framework is as of yet
known. Work is currently in progress on this very important issue.

It is tempting to try to apply the new class of gravity theories
to the fundamental problems of cosmology, namely those of
``dark matter'' and ``dark energy''. We have indicated how
this might be possible, but much more work is needed, in particular
on coupling of these theories to massive matter fields, to
see if any realistic model can be built along these lines.

We have reviewed a conjecture to the effect that the described
class of theories (with varying $\phi$) is closed under the
renormalization. Some arguments in support of this conjecture
were given, but its status remains open.

\section*{Acknowledgments} The author is grateful to Yuri Shtanov for collaboration on various
aspects of non-metric gravity theories and to Jean-Marc Schlenker and Sergei Winitzki for
stimulating discussions. The author is supported by the Advanced EPSRC Fellowship.

\end{document}